\documentclass[aps,10pt, pre,reprint,superscriptaddress,showpacs]{revtex4-1}
\usepackage{graphicx}
\usepackage{amsmath}
\usepackage{soul,xcolor}
\usepackage{color}
\usepackage{placeins}
\usepackage{braket}
\usepackage[normalem]{ulem}

\usepackage{mathtools}

\definecolor{dgreen}{rgb}{0.0, 0.5, 0.0}

\newcommand{\veck}{\mathbf k}

\newcommand{\vecp}{\mathbf p}
\newcommand{\vecq}{\mathbf q}

\newcommand{\vecr}{\mathbf r}

\newcommand{\vecR}{\mathbf R}

\newcommand{\volume}{\mathcal{V}}

\newcommand{\ad}{\hat a^\dagger}
\renewcommand{\a}{\hat a}

\begin{document}

\preprint{APS/123-QED}

\title{Creation of  Rydberg Polarons in a Bose Gas}
\author{F. Camargo}
\affiliation{
 Department of Physics \& Astronomy, Rice University, Houston, TX 77251, USA
}
\author{R. Schmidt }
\affiliation{
ITAMP, Harvard-Smithsonian Center for Astrophysics, Cambridge, MA 02138, USA
}%
\affiliation{
Department of Physics, Harvard University, Cambridge, MA 02138, USA
}%
\affiliation{
 Institute of Quantum Electronics, ETH Z\"urich, CH-8093 Z\"urich, Switzerland
}

\author{J.\,D. Whalen}
\author{R. Ding}
\affiliation{
 Department of Physics \& Astronomy, Rice University, Houston, TX 77251, USA
}
\author{G. Woehl Jr.}
\affiliation{
 Department of Physics \& Astronomy, Rice University, Houston, TX 77251, USA
}
\affiliation{
 Instituto de Estudos Avan\c{c}ados, 12.228-001 S\~{a}o Jos\'{e} dos Campos, S\~{a}o Paulo, Brazil
}

\author{S. Yoshida} 
\author{J. Burgd\"orfer}
\affiliation{%
Institute for Theoretical Physics, Vienna University of Technology, Vienna, Austria, EU
}%

\author{F.\,B. Dunning}
\affiliation{
 Department of Physics \& Astronomy, Rice University, Houston, TX 77251, USA
}

\author{H.\,R. Sadeghpour }
\affiliation{
ITAMP, Harvard-Smithsonian Center for Astrophysics, Cambridge, MA 02138, USA
}%

\author{E. Demler }
\affiliation{
Department of Physics, Harvard University, Cambridge, MA 02138, USA
}%

\author{T.\,C. Killian}%
\affiliation{
 Department of Physics \& Astronomy, Rice University, Houston, TX 77251, USA
}

\date{\today}

\begin{abstract}
{We report spectroscopic observation of Rydberg polarons in an atomic Bose gas. Polarons are created by excitation of Rydberg atoms {as impurities} in a strontium Bose-Einstein condensate. {They} are distinguished from  previously studied polarons  by {macroscopic} occupation of bound molecular states that arise from scattering  {of the weakly}  bound Rydberg electron  {from} ground-state atoms.  The absence of a $p$-wave resonance in the low-energy electron-atom scattering in Sr introduces a universal behavior in the Rydberg spectral lineshape
{and in scaling of the spectral width (narrowing) with the Rydberg principal quantum number, $n$}.
 Spectral
 features are  described with a functional determinant approach (FDA) that solves an extended Fr\"{o}hlich Hamiltonian for {a mobile} impurity in a Bose gas. {Excited states of polyatomic Rydberg molecules (trimers, tetrameters, and pentamers) are experimentally resolved  and accurately reproduced with FDA}.}

\end{abstract}

\maketitle
The interaction of an impurity with a deformable medium can lead to a collective response and formation of quasi-particles known as polarons, consisting of the impurity dressed by excitations of the background medium. Polarons play important roles in conduction in ionic crystals and polar semiconductors \cite{dpe82}, spin-current transport in organic semiconductors \cite{wak14}, optical absorption of two-dimensional materials \cite{Schmidt2012b,Sidler2016,MacDonnald2017}, and  collective excitations in strongly interacting fermionic and bosonic ultracold gases \cite{swa09,hvk16,jws16}. Here, we report spectroscopic observation of  {Rydberg} polarons formed through excitation of Sr  Rydberg atoms in a strontium  Bose-Einstein condensate (BEC), which represent a new class of impurity states beyond those typically seen in condensed matter settings \cite{dal09}.
 Rydberg polarons are distinguished by {macroscopic  occupation of bound molecular states that arise from interaction of ground state atoms with the Rydberg electron.}
  A functional determinant approach (FDA) \cite{ssd16}, {adapted to include effects of impurity recoil, } solves an extended Fr\"{o}hlich Hamiltonian for  {an} impurity in a Bose gas { and} {accurately reproduces the observed excitation spectrum. }

Rydberg-excitation experiments have been carried out previously using a Rb BEC \cite{bkg13,sln16}. In Rb, however,
a  $p$-wave shape resonance \cite{fab86,ckf02,hgs02} for e-Rb scattering introduces highly non-universal density  and $n$ dependence of Rydberg absorption spectra,  complicating the identification of polaron features.
{In Sr, such a shape resonance is absent}.  {Deviations in the Sr excitation spectrum  from a mean-field prediction reveal
 that in the s-wave-dominated regime,  the shape of the  polaron spectrum and the scaling of the spectral width with $n$ are {insensitive to quantitative details of the underlying potential, and, in this sense, universal}. Accessing this window of universality is new in Rydberg physics.

{The formation of
Rydberg  and typical condensed-matter polarons \cite{dal09}
can both be understood as
dressing of the impurity by excitations of the bosonic bath, which entangles the impurity momentum with that of the bath excitations \cite{SuppMaterial},}
but there are important differences. Most importantly,  the spectral width for a region of uniform density shows that Rydberg polarons exist not as the ground state of the many-body system but   rather as a set of excited states, which is unusual in polaron physics.
 {This highlights the special nature of Rydberg polaron formation where no single many-body state dominates the non-equilibrium dynamics at late times.}
 {Furthermore, the} description of Rydberg polarons  relies on multiple excitations from the BEC - on the order of the number of atoms within the Rydberg orbit, while other polaron states are {often}   well described by  including only single-excitation terms. In addition, whereas most polarons in condensed matter systems can be described fully with
 {excitations to unbound states such as free particle or phonon modes,}
  for Rydberg polarons it is essential to  include negative-energy bound states.

  {Bound} states are also important for Fermi 
  \cite{swa09,kohstall2012,koschorreck2012,Massignan2014,Cetina96,Schmidt2017} and
  Bose polarons \cite{hvk16,jws16, PhysRevLett.117.113002,PhysRevLett.115.160401, PhysRevLett.115.125302, PhysRevA.92.033612, PhysRevA.92.023623,  PhysRevA.94.063640,  PhysRevA.90.013618} consisting of ground-state impurity atoms interacting with atoms in a background Fermi gas or BEC, respectively. In these cases,   contact interactions are tuned with a Feshbach resonance, while  Rydberg impurities introduce  long-range interactions with strength tuned by changing \textit{n}.

The interaction between a Rydberg atom and a bosonic medium can be described
with the
Born-Oppenheimer potential for a ground-state atom a distance $\vecr$ from the Rydberg impurity  \cite{fer34,gds00,fab86,ckf02,omo77}
\begin{eqnarray}
V(\vecr)&=&\frac{2\pi \hbar^2}{m_e} A_s|\Psi(\mathbf{r})|^2
 +\frac{6\pi \hbar^2 }{m_e} A_p^3|\overrightarrow{\nabla}\Psi(\mathbf{r})|^2,   \label{RydbergInteraction}
\end{eqnarray}
where $\Psi(\mathbf{r})$ is the Rydberg electron wave function, $A_s$ and $A_p$ are the momentum-dependent $s$-wave and $p$-wave scattering lengths, and $m_e$ is the electron mass.
When $A_s<0$, $V(\mathbf{r})$ can support molecular states  with one or more ground-state atoms bound to the impurity \cite{gds00,bbn09,dad15}. The typical form of such a potential is shown in Fig.~\ref{Fig:SchematicandFewBodyDataAndTheory}, which illustrates schematically the formation of a Rydberg polaron. Laser excitation of an atom in the BEC creates a Rydberg impurity and projects the system into a superposition
of bound and scattering states $\ket{\beta_i}$. {While} bound states are confined in the potential $V({\mathbf{r}})$,  scattering states can encompass the entire BEC.

\begin{figure}
\includegraphics[clip=true,keepaspectratio=true,width=3in,trim=0in 0in 0in 0in]{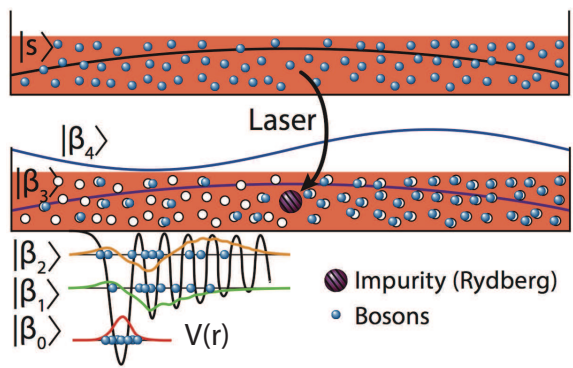}
\caption{(Color online) Schematic of the excitation of a Rydberg polaron in a uniform-density BEC. Laser excitation projects the system into final configurations involving atoms in bound and scattering states $\ket{\beta_i}$.
Bound states are confined in the Rydberg potential $V(\textbf{r})$.
\label{Fig:SchematicandFewBodyDataAndTheory}}
\end{figure}

The Hamiltonian for a mobile impurity with mass $M$ at position $\hat \vecR$ and momentum ${\hat \vecp}$, interacting with a gas of bosons is
\begin{eqnarray}\label{appHFullFirstQuant}
\hat H= \frac{\hat \vecp^2}{2M}
+\sum_\veck \epsilon_\veck \ad_\veck\a_\veck
+\frac{1}{\volume}\sum_{ \veck\vecq}V(\vecq)e^{{-}i\vecq \hat \vecR}\ad_{\veck+\vecq}\a_{\veck},
\end{eqnarray}
where $\volume$ is the quantization volume, and $\a_\veck$ ($\ad_\veck$) are the annihilation (creation) operators for the bosons with dispersion $\epsilon_\veck=\frac{\hbar^2\veck^2}{2m}$ (for details see \cite{SuppMaterial}). The last term, which gives rise to   {polaron formation}, describes the Rydberg impurity-boson interaction where $V(\vecq)$ is the Fourier transform of    {Eq.}~\eqref{RydbergInteraction}. In our experiments, the impurity-boson interaction is much stronger than the interaction between bosons, which    { hence can be} omitted {when calculating  the excitation spectrum from a region of constant density}.

 {The} Hamiltonian  {of} Eq.\ {\eqref{appHFullFirstQuant}}     {is the basis for} various polaron models \cite{SuppMaterial}. For instance, applying the  Bogoliubov approximation and keeping only  {terms} that are linear in bosonic field operators yields the widely-studied Fr\"ohlich Hamiltonian \cite{tco09}. However, the Fr\"ohlich Hamiltonian does not account correctly for scattering, including formation of bound states \cite{rsc13, PhysRevLett.115.160401, PhysRevLett.115.125302, PhysRevA.92.033612, PhysRevA.92.023623, PhysRevLett.117.113002, PhysRevA.94.063640, jws16, hvk16, PhysRevA.90.013618,Ashida2017}. {Retaining all impurity-boson interaction terms   yields an extended Fr\"ohlich Hamiltonian, which is required (and used here) to { account for}  the role that bound molecular states play in Rydberg polaron formation.

The Rydberg excitation spectrum in  {the} linear response   is given by $A(\nu)\propto\sum_\gamma \left|\Braket{\Psi_\gamma | {\Psi_{\rm BEC}}}\right|^2 \delta[\nu-(E_\gamma-{E_{\rm BEC}})/h]$, for laser detuning $\nu$. Here {$\ket{\Psi_{\rm BEC}}$} is the BEC ground state {before excitation of the impurity,} with energy $E_{\rm BEC}$, and the sum is over all many-body eigenstates $\Ket{{\Psi_\gamma}}$ of Eq.~\eqref{appHFullFirstQuant} {describing the system after impurity excitation}, with energy $E_\gamma$ \cite{SuppMaterial}. The states $\Ket{{\Psi_\gamma}}$ are constructed by the occupation of single-particle bound and scattering
states $\ket{\beta_i}$, which are modified by the Rydberg potential (see Fig.~\ref{Fig:SchematicandFewBodyDataAndTheory}). $A(\nu)$ is evaluated with a FDA that was previously developed to describe only the case of immobile impurities \cite{ssd16}.
{Here, we extend the previous work with
a canonical transformation that transforms the system into the frame co-moving with
the impurity \cite{SuppMaterial,Lee1953,Girardeau1961} and
decouples the impurity and boson Hilbert spaces, mapping the problem onto a purely bosonic system \cite{PhysRevLett.117.113002}. This allows treatment of impurity motion, which is crucial for an accurate description of the observed spectra.}
{Spectra are obtained} from the unitary {time} evolution of the Hamiltonian in Eq.~\eqref{appHFullFirstQuant} including all orders of the impurity-Bose interaction. {Finally, we} average the spectrum over
the atomic density profile \cite{SuppMaterial}.}

The Rydberg excitations are carried out in a $^{84}$Sr BEC in a crossed-beam, optical-dipole trap \cite{mmy09,sth09}. The BEC is close to spherically symmetric, with $3.5\times 10^5$\,atoms, a peak condensate density $\rho_{\textup{max}}=3.6\times10^{14}\,\textup{cm}^{-3}$, and a Thomas-Fermi radius $R_{TF}=8\,\mu$m. The  condensate fraction is $\eta \sim75\%$ and   the temperature is $T\sim 150$\,nK. These parameters are determined from time-of-flight measurements using resonant absorption imaging on the $^{1}S_0\rightarrow{^{1}P_1}$ transition with 461\,nm light. The average nearest-neighbor atom separation at $\rho_{\textup{max}}$ is 80\,nm.

Rydberg spectra are obtained by driving the $5s^2~^{1}S_0\rightarrow5s5p~^{3}P_1\rightarrow5sns~^{3}S_1$ two-photon transition using 689\,nm and 319\,nm light \cite{dad15,cwd16}. The 689\,nm light is blue-detuned from the intermediate state by 80\,MHz while the frequency of the 319\,nm laser is scanned. The laser excitation is applied in a 2\,$\mu$s pulse,  immediately followed by an electric-field ramp that ionizes Rydberg atoms and molecules in about 1\,$\mu$s   and directs the product electrons towards a micro-channel-plate detector. The time from excitation to detection is kept short to avoid density-dependent Rydberg-atom loss \cite{cwd16,sle16}, which would distort the spectrum. Typical intensities of the excitation beams are $200\,\textup{mW/cm}^{2}$ and $6\,\textup{W/cm}^{2}$ for the 689\,nm and 319\,nm beams, respectively, and they are uniform across the BEC. This results in a Rydberg excitation rate well below one per pulse, avoiding Rydberg-blockade effects and ensuring a linear-response measurement.
Excitation and detection are repeated at a 4\,kHz rate. For each frequency point in the spectrum, a new trapped sample is used and the signal from 1000 pulses is acquired. The peak atom density varies by less than $5\%$ for each point.

\begin{figure}
\includegraphics[clip=true,keepaspectratio=true,width=3.25in,trim=0in 0in 0in 0in]{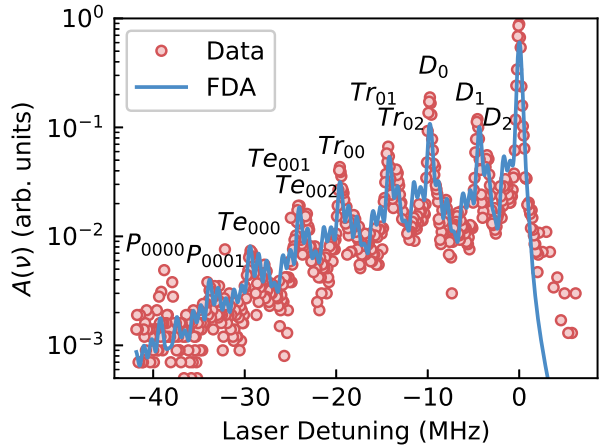}
\caption{(Color online) Spectrum  for excitation of the $5s38s~{^{3}S_1}$ Rydberg state in the few-body regime. Pure atomic excitation is at zero detuning, and lines for dimers ($\textup{X}=\textup{D}$) through pentamers ($\textup{X}=\textup{P}$) are visible. Subscripts (X$_{i j ...}$) indicate the  molecular levels of the Rydberg potential occupied by atoms after laser excitation.  {FDA}
matches the data over several orders of magnitude in signal strength, which is normalized to the peak at zero detuning.
\label{Fig:FewBodySpectrum}}
\end{figure}

At low principal quantum numbers, resolved Rydberg-molecular lines dominate the spectrum and  {are well understood} \cite{gkb14}. This is shown for Sr$(5s38s)$ excitation in Fig.~\ref{Fig:FewBodySpectrum}, for which on average  three atoms are within the Rydberg orbital ($N_{\textup{orb}}= \rho_{\textup{max}}4\pi r_R^3/3$), where $r_R$ is the radius of the outer lobe of the wave function.
Individual spectral lines correspond to Rydberg molecular states with one (\underline{D}imer), two (\underline{Tr}imer), three (\underline{Te}tramer), and four (\underline{P}entamer) bath atoms  bound in the available levels. The intricate  structure of the molecular lines is reminiscent of the nuclear shell model.

We find remarkable agreement between theory and experiment over several orders of magnitude of spectral strength. For theoretical calculations, the sample temperature, total atom number, trap frequencies, and overall signal intensity are taken as adjustable parameters and lie within experimental uncertainty \cite{SuppMaterial}.

\begin{figure}
\includegraphics[clip=true,keepaspectratio=true,width=3.0in,trim=0in 0in 0in 0in]{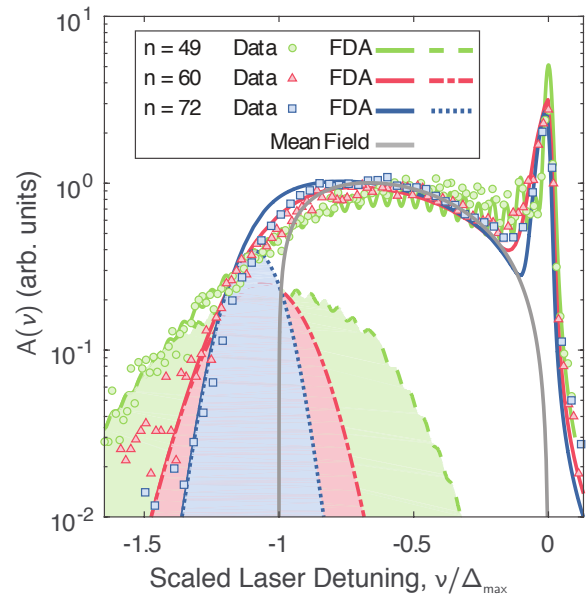}
\caption{(Color online) Rydberg  excitation in the many-body regime for $n=49, 60, 72$.  The mean-field prediction is for the condensate atoms only.  Deviations between the mean-field result and data at small detuning ($\nu/\Delta_{\textup{max}}<0.3$) represent contributions from the non-condensed thermal atoms.   {The shaded regions represent the predicted polaron response from the center of the atom cloud as obtained from FDA.}   Theory for  {this} high-density region matches the tail of the experimental data and represents the Rydberg-impurity spectral function for a nearly uniform-density region. For uniform density, the width of the spectrum scales as $n^{-3}$.
\label{Fig:MolSpecLogScaled}}
\end{figure}

For higher  {n}, the average number of ground-state atoms per Rydberg-orbital volume increases
to $13$, $50,$ and $160$ atoms for $n=49,\,60$ and $72$, respectively, and many-body effects become relevant (Fig.~\ref{Fig:MolSpecLogScaled}). The   interplay of few- and many-body physics is particularly evident for $n=49$. Here, a series of up to seven-fold-occupied molecular bound states  can be identified for scaled detuning between zero and -0.5. With increasing $n$, the spectrum evolves into a broad continuum, which is the signature of many-body dressing of the impurity.
 {The peak in spectral support  at zero detuning, comes almost exclusively from the low-density, non-condensed atoms.}  The FDA, which  accounts for quantum many-body effects, describes the data well (details in \cite{SuppMaterial}).

A   mean-field treatment using the local density approximation  helps to elucidate the spectral signatures of polaron excitation in the regime of large numbers of atoms per Rydberg orbital. This treatment is reminiscent of Fermi's original description of Rydberg excitation in dense thermal gases \cite{gkb14,fer34} and predicts an excitation spectrum of the form $A(\nu)\propto \int  d^3r\, \rho(\mathbf{r}) \delta[\nu-\Delta(\mathbf{r})]$. An atom in a spatial region with density $\rho(\mathbf{r})$ contributes to the spectrum at a detuning from unperturbed atomic resonance of $\Delta(\mathbf{r}) =\rho(\mathbf{r}) \int  d^3r'\, V(\mathbf{r}-\mathbf{r}')/h$. This mean-field shift   reflects a spatial average of the interaction between Rydberg and ground-state atoms ({cf. Eq.}~\eqref{RydbergInteraction}). At fixed density, $\Delta$ varies slightly with $n$ \cite{SuppMaterial}.

In Fig.~\ref{Fig:MolSpecLogScaled}, the frequency axis is scaled by the fit value of the mean-field shift for the peak density in the BEC ($-\Delta_{\textup{max}}$) \cite{SuppMaterial}, and  {the} {Rydberg-excitation}   intensity is  normalized  {to its} integral neglecting  {the thermal} contribution. With this scaling, the mean-field prediction is   {independent of  {n}. The rough agreement between experiment and the mean field prediction implies that the shape of the full experimental spectrum is dominated by broadening due to the inhomogeneous, in-trap density distribution.  {While the mean-field matches the response at intermediate frequencies, it}   {fails   to describe the resolved few-body features   and} the red-detuned tails of the spectra in Fig.~\ref{Fig:MolSpecLogScaled}.

 It is the deviations from mean-field theory that provide key insight into the nature of Rydberg polarons. While polarons in condensed matter systems [6] are often described through dressing by long wave-length excitations of the medium such as phonons, the formation of Rydberg polarons relies on the dressing by localized states bound to the impurity with discrete, negative energies. These molecular levels are revealed  {as} resolved features at small detuning for $n = 49$. Massive, many-body occupation of these bound states gives rise to the formation of Rydberg polarons and the broad tail at far red detuning.

{ FDA predicts that Rydberg polaron formation  leads to a Gaussian profile of the excitation spectrum for systems of uniform density, with a width that decreases with increasing $n$, which are both beyond-mean-field effects.
 To investigate these predictions, we focus on the spectral tail at large red detuning where a clear variation between the different data sets is found.  Using FDA, we are able to isolate the spectral contribution from a small, central region with the highest density (shaded regions in Fig.~\ref{Fig:MolSpecLogScaled}). Comparison between experiment and FDA shows that the tail arises entirely from this high-density region.  In this region the average density is $\langle \rho \rangle=0.92 \rho_{\rm max}$. The small standard deviation ($\sqrt{\langle (\Delta \rho)^2 \rangle} /\langle \rho \rangle=0.04$) shows that the  spectral width reflected in the tail cannot be explained by  density inhomogeneity. Since the observed spectral width likewise cannot be explained by laser resolution (400\,kHz),  we attribute it to the intrinsic excitation of Rydberg polarons.  The  agreement between the simulated Rydberg polaron response and the observed data over an order of magnitude variation in the signal strength confirms the FDA predictions.}

The emergence of a Gaussian profile is {one of} the defining properties of Rydberg polaron formation. Its presence can be understood by decomposing the spectral  function $A(\nu)$ in terms of excitations from the BEC ground state to interacting single-particle states. The resulting multinomial distribution is dominated by a few terms with large probability $p_i=\left|\Braket{\beta_i|s}\right|^{2}$ for bath particles in state $\ket{s}$ to scatter, via interaction with the Rydberg atom, into states $\ket{\beta_i}$ (see Fig.~\ref{Fig:SchematicandFewBodyDataAndTheory}).
For  {a model with} $N$  {particles} and only two interacting single-particle states, {i.e.} one bound state ({e.g.} $\ket{\beta_0}$, Fig.~\ref{Fig:SchematicandFewBodyDataAndTheory}), with energy $\epsilon_B$
  and overlap $p_0 \ll 1$, and one low-energy scattering state (e.g. $|\beta_s\rangle\equiv |\beta_3\rangle$) with energy $\epsilon_3\simeq 0$ and   {$p_3 = 1 - p_0  $}, a binomial distribution results,
\begin{equation}\label{binom}
A(\nu)=\sum_{j=1}^N  \binom {N}{j} p_0^j (1-p_0)^{N-j}\delta(\nu-j\epsilon_B/h).
\end{equation}
For   { large $N$ Eq.}~\eqref{binom} becomes a Gaussian with mean energy $\mu=Np_0\epsilon_B$ and variance $Np_0(1-p_0)\,\epsilon_B^2$. Since the single-particle overlap of $\ket{s}$ with the bound state $\ket{\beta_0}$ is small, the Gaussian width becomes $\sqrt{Np_0}\,\epsilon_B$. We identify $Np_0 \equiv N_{\textup{orb}}= 4\pi \rho r_R^3/3$, where $r_R\sim n^2$ is the Rydberg electron radius, and assume that the Gaussian mean energy corresponds to the mean-field energy $h\Delta$. The Gaussian width then varies  as $\sim \sqrt{\rho}/n^3$, in agreement with experiment and numerical simulation of the full Hamiltonian (Fig.~\ref{Fig:MolSpecLogScaled}). {The spectrum is sensitive to density-density correlations in the background gas, which display Poissonian fluctuations for a BEC.} The decreasing width with decreasing density also explains why the data and   mean-field prediction agree well at small detuning after subtracting the contribution from the thermal fraction. The Gaussian lineshape and $\sim \sqrt{\rho}/n^3$ scaling are  universal features that emerge when $s$-wave electron-atom scattering dominates, and they are masked by sensitivity to quantitative details of the underlying interaction in the presence of a shape resonance,  as in Rb \cite{sln16}.


The spectrum $A(\nu)$ in the approximate expression (3)
consists of delta-shaped peaks corresponding to
quantized molecular levels. {In the accurate FDA description},   the binomial distribution is {effectively } replaced by a
multinomial distribution that yields denser energy levels and includes coupling to scattering states, which, in principle, broadens molecular levels asymmetrically to positive
energies.
 But this effect is  too small to be resolved
within our experimental resolution.}
The experimentally observed broadening to the blue of unperturbed atomic resonance, most clearly visible in  Fig.\ \ref{Fig:FewBodySpectrum}, does not arise from known mechanisms connected to the presence of a shape resonance \cite{sln16,trb12}, {nor can it be attributed to finite temperature}, and it remains a subject for future studies.

When the spacings of molecular levels become comparable to the broadening due to scattering states {or experimental resolution}, the
spectrum becomes a continuous distribution. In this limit the spectrum can also be well described
 by a classical statistical model  as, e.g., used  to describe Rydberg impurity excitations in a Rb condensate \cite{sln16}.
 As shown in \cite{SuppMaterial}, this model corresponds to an approximation in which  the motions of the impurity and bath atoms are neglected.    Since this model `freezes' all dynamics in the system it cannot describe Rydberg molecules (visible as resolved lines in the spectrum), which are the underlying building blocks of Rydberg polarons.


In conclusion, we present experimental evidence  for the creation of Rydberg polarons consisting of an impurity Rydberg atom dressed by a large number of bound   excitations of Bose-condensed Sr atoms. {By focusing on the red-detuned spectral tail, we show that a gaussian lineshape and $1/n^3$ scaling of the linewidth with principal quantum number are universal features of the Rydberg-polaron excitation spectrum in a BEC when s-wave electron-atom interactions dominate. The spectral width reflects that Rydberg polarons  are created in excited states of the impurity-bath system  that originate from the interplay of many-body dressing and few-body molecule formation. The spectral width can serve as a local probe of density-density correlations in the BEC.} Our observations are in remarkable agreement with FDA predictions including impurity recoil,  emphasizing the quantum nature of the polaron.
Important questions remain to be explored, such as whether the interactions are strong enough to lead to self-localization of the impurity, and
{the extent to which notions of effective mass, mobility, and polaron-polaron interactions are applicable to Rydberg polarons.}
 { Ramsey interferometry experiments
can probe the intrinsic quantum nature of the dressing of Rydberg impurities in real time \cite{PhysRevLett.115.135302, PhysRevX.2.041020, Cetina96,Schmidt2017}.}

Research supported by the AFOSR (FA9550-14-1-0007), the NSF (1301773, 1600059, and 1205946), the Robert A, Welch Foundation (C-0734 and C-1844), the FWF(Austria) (P23359-N16, and FWF-SFB049 NextLite).  The Vienna scientific cluster was used for the calculations.
R.~S. and H.~R.~S. were supported by the NSF through a grant for the Institute for Theoretical Atomic, Molecular, and Optical Physics at Harvard University and the Smithsonian Astrophysical Observatory. R.~S. acknowledges support from the ETH Pauli Center for Theoretical Studies.
E.~D. acknowledges support from Harvard-MIT CUA, NSF Grant (DMR-1308435), AFOSR Quantum Simulation MURI, the ARO-MURI on Atomtronics, and support from  Dr.~Max R\"ossler, the Walter Haefner Foundation and the ETH Foundation. T.~C.~K acknowledges support from Yale University during writing of this manuscript.



%

\end{document}